\documentclass[prd,superscriptaddress,showpacs,preprintnumbers,amsmath,amssymb]{revtex4}
\usepackage{graphicx}% Include figure files
\usepackage{dcolumn}% Align table columns on decimal point
\usepackage{bm}% bold math
\usepackage{epsfig}
\usepackage{amsfonts}

\begin{document}

\title{Some properties of accelerating observers in the Schwarzschild space}

\author{Ya-Peng Hu}
  \email{huzhengzhong2050@163.com}
  \affiliation{Institute of Theoretical Physics, Chinese Academy of Sciences, P.O.
Box 2735, Beijing 100190, China}
 \affiliation{Graduate School of the Chinese Academy of Sciences, Beijing 100039,
China}
\author{Gui-Hua Tian}
   \email{hua2007@126.com}
   \affiliation{School of Science, Beijing University Of
   Posts and Telecommunications, Beijing 100876, China}
\author{Zheng Zhao}
   \email{zhaoz43@hotmail.com}
   \affiliation {Department of Physics,
Beijing Normal University, Beijing 100875, China}

\begin{abstract}
It is well known that observers will be accelerated when they
approach the planets. Thus, discussing the properties of
accelerating observers in the Schwarzschild space is of sense. For
the sake of simplicity, we can construct these observers' world
lines by comparing with the observers in the Hawking effect and
Unruh effect, whose world lines are both hyperbolic curves in the
appropriate coordinates. We do it after defining a certain special
null hypersurface in the Kruskal coordinates. Our result shows that
these accelerating observers defined in our paper can also detect
the radiation with the analogy of the Unruh effect, though locally.
Furthermore, we conjecture that, for any null hypersurface in any
spacetime, the corresponding observers who can at least locally
detect the radiation from it can be found.

\end{abstract}

 \maketitle

\section{Introduction}
Simply viewed from the astronomical exploration, it's of sense to
theoretically consider the properties of accelerating observers when
approach the planets. However, the spacetime is an unity in general
relativity \cite{1}, and the concept of accelerating in the universe
is ambiguous. Since the outside spacetime of the planet can always
be approximately viewed as the Schwarzschild spacetime, we can
naturally define the concept of acceleration in terms of the second
time derivatives of position in the usual Schwarzschild space with
the killing vector$(\frac{\partial }{\partial t})^{a}$.

 On the other hand, there have been many works considering the properties
 of accelerating observers in various of different spacetimes and conditions \cite{2,3,4,5,6,7,8,9,10,11,12,13,14}.
 And two significant representations of these works are the discovery of
Hawking effect and Unruh effect\cite{2,5}. These two effects both
conclude that the accelerating observers can detect the radiation as
they are in the thermal bath. Moreover, the observers' accelerations
are uniform in the spacetime. In addition, both effects have another
common property of their observers, which is that the observers'
world lines are all hyperbolic in the appropriate coordinates. Thus,
for simplicity, we can construct our accelerating observers' world
lines in the Schwarzschild space by taking the analogy of these two
effects. In the Kruskal coordinates, after defining a certain
special null hypersurface, we can define the world lines of our
accelerating observers to be the hyperbolic curves of which the
asymptote is just the null hypersurface itself. And in the latter we
can prove that the obsevers defined in this way truly accelerate in
the Schwarzschild space. It should be emphasized that there are
other constructions of the observers accelerating in the
Schwarzschild space, but the properties of the observers may be
different. For example, for the free drop observer and the static
observer in the Schwarzschild space, the former corresponds to a
geodesic observer and detects no radiation, while the latter can
detect the well-known Hawking effect. In fact, the Hawking effect
with the static observer can be viewed as a special case in our
discussion. And with the very similar trick (defining the null
hypersurface first), the authors in Ref[15] discussed various of
properties of accelerating observers in the Minkowski spacetime.

The present work starts by defining a certain null hypersurface and
its corresponding observers in the Kruskal coordinates. After
discussing some properties of these observers, we will show that the
event horizon with the static observers in the Hawking effect can be
viewed as the special case in our discussion.

Then, another more significant property of these observers,
detecting radiation, is discussed in section 3 by using the Damour-
Ruffini method\cite{13}. The radiation can be viewed as the analogy
of the Unruh effect. However, because $(\frac{\partial }{\partial
\eta})^{a}$ is not a killing vector, the thermal spectrum would be
just local. That is, it is detected just by the observers near the
horizon and it can't be detected consistently with the observers in
the infinity.

Finally, in section 4, we will first show that these observers
defined in this way truly accelerate in the Schwarzschild space, and
then we give a brief conclusion and discussion. Note that, although
the effective temperature is also very low to detect in the
experiment as those in many other similar works, the theoretical
discussion is still of sense. Particularly, we conjecture that, for
any null hypersurface in any spacetime, the corresponding observers
who can at least locally detect the radiation from it can be found.
This conjecture can be seen to be partly supported in the Ref[15].
And its application can also be seen in Ref [16] where the authors
defined the local temperature with this conjecture and obtained some
very interesting results.

\section{The null hypersurface and its corresponding observers by definition}

The line element of the Schwarzschild spacetime in the Kruskal
coordinates is given by \cite{1}

\begin{equation}
ds^{2}=\frac{32M^{3}}{r}e^{-\frac{r}{2M}}(-dT^{2}+dX^{2})+r^{2}(d\theta
^{2}+\sin ^{2}\theta d\varphi ^{2}).  \label{1}
\end{equation}%
where

\begin{equation}
(\frac{r}{2M}-1)e^{\frac{r}{2M}}=X^{2}-T^{2}.  \label{2}
\end{equation}

In the following, we restrict our discussions only in the region
$r>2M$. Thus, we first define the null hypersurface in this region

\begin{equation}
T=\pm (X-X_{0}).  \label{3}
\end{equation}%
where $X_{0}$ is a positive constant. It's easily seen that, when
$X_{0}=0$, the null hypersurface defined in (3) can become the usual
event horizon of the Schwarzschild spacetime.

According to (3), the corresponding observer is

\begin{equation}
(X-X_{0})^{2}-T^{2}=e^{2a\xi _{0}},\theta =\theta _{0},\varphi =\varphi _{0}.
\label{4}
\end{equation}%
where $a,\xi _{0},\theta _{0},\varphi _{0}$ are also constants. In
fact, a family of observers that we are interested in can be defined
by choosing different $\xi _{0}$. It's obvious that all the world
lines of these observers are hyperbolic curves in the Kruskal
coordinates. Furthermore, the null hypersurface defined in (3) can
be viewed as their horizon, which can be easily obtained from the
T-X diagram in Figure 1.

In fact, by these observers we can define a new coordinate system
$\{\eta ,\xi ,\theta ,\varphi \}$ that

\begin{equation}
X=X_{0}+e^{a\xi }cha\eta ,T=e^{a\xi }sha\eta .  \label{5}
\end{equation}%
And the metric in this coordinates is

\begin{eqnarray}
ds^{2} &=&\frac{32M^{3}}{r}e^{-\frac{r}{2M}}e^{2a\xi }a^{2}(-d\eta ^{2}+d\xi
^{2})+r^{2}(d\theta ^{2}+\sin ^{2}\theta d\varphi ^{2})  \notag \\
&=&g_{00}d\eta ^{2}+g_{11}d\xi ^{2}+r^{2}(d\theta ^{2}+\sin ^{2}\theta
d\varphi ^{2}).  \label{6}
\end{eqnarray}%
where

\begin{equation}
(\frac{r}{2M}-1)e^{\frac{r}{2M}}=X_{0}^{2}+2e^{a\xi }X_{0}ch(a\eta
)+e^{2a\xi }.  \label{7}
\end{equation}%
In this co-moving coordinate system, it's obvious that observers
labelled by different $\xi$ are at rest, and the horizon is at
$\xi =-\infty $, which is just the null hypersurface defined in
(3) and agrees with the former conclusion from Fig 1.

The tangent vector of the observer's worldline ($\xi =\xi _{0}$)
defined in (5) is

\begin{equation}
(\frac{\partial }{\partial \eta })^{a}=\frac{\partial x^{\mu }}{\partial
\eta }(\frac{\partial }{\partial x^{\mu }})^{a}=a[T(\frac{\partial }{%
\partial X})^{a}+(X-X_{0})(\frac{\partial }{\partial T})^{a}].  \label{8}
\end{equation}%
Re-parameterize the world line with the proper time $\tau $, we
get

\begin{equation}
(\frac{\partial }{\partial \eta })^{a}=\frac{d\tau }{d\eta }(\frac{\partial
}{\partial \tau })^{a}=\gamma (\frac{\partial }{\partial \tau })^{a},\gamma
\equiv \frac{d\tau }{d\eta }.  \label{9}
\end{equation}%
where $(\frac{\partial }{\partial \tau })^{a}$ is the 4-velocity.
Using the unitary property of the 4-velocity and the metric, we
find

\begin{eqnarray}
\gamma ^{2} &\equiv &(\frac{d\tau }{d\eta })^{2}=a^{2}e^{2a\xi _{0}}\frac{%
32M^{3}}{r}e^{-\frac{r}{2M}},  \notag \\
\Gamma _{00}^{0} &=&\Gamma _{11}^{0}=\Gamma _{01}^{1}=\frac{2(r+2M)MT}{%
r^{2}e^{\frac{r}{2M}}},  \notag \\
\Gamma _{01}^{0} &=&\Gamma _{00}^{1}=\Gamma _{11}^{1}=-\frac{2(r+2M)MX}{%
r^{2}e^{\frac{r}{2M}}}.  \label{10}
\end{eqnarray}%
where the nonzero Christoffel coefficients are correlated with the
Kruskal coordinates. Thus, the 4-velocity and the 4-acceleration are

\begin{eqnarray}
T^{a} &=&(\frac{\partial }{\partial \tau })^{a}=\frac{1}{\gamma }(\frac{%
\partial }{\partial \eta })^{a}=\frac{a}{\gamma }[T(\frac{\partial }{%
\partial X})^{a}+(X-X_{0})(\frac{\partial }{\partial T})^{a}]  \notag \\
A^{b} &=&\frac{e^{-2a\xi }}{\frac{32M^{3}}{r}e^{-\frac{r}{2M}}}B[T(\frac{%
\partial }{\partial T})^{b}+(X-X_{0})(\frac{\partial }{\partial X})^{b}].
\label{11}
\end{eqnarray}%
where

\begin{equation}
B\equiv 1-[(X-X_{0})X_{0}+e^{2a\xi _{0}}]\frac{2M(r+2M)}{r^{2}}e^{-\frac{r}{%
2M}}.  \label{12}
\end{equation}%
And the proper acceleration can also be obtained

\begin{equation}
A\equiv \left\vert A^{a}\right\vert =\frac{1}{e^{a\xi _{0}}}(\frac{32M^{3}}{r%
}e^{-\frac{r}{2M}})^{-1/2}B.  \label{13}
\end{equation}

From (8) to (13), we can draw a brief conclusion about the
properties of these observers. First, these observers' world lines
are truly timelike such that they can be considered as detectors.
Second, the proper acceleration changes along with observer's proper
time or $\eta$, which is different from that of a static observer in
the Schwarzschild space-time, which is a constant. Third, when the
observer get closer to the horizon or in the limit $\xi
_{0}\rightarrow -\infty $, the proper acceleration becomes positive
infinite, which is a general property of the horizon. And this can
also be understood that the proper acceleration of the light may be
positive infinite (more details can be seen in the Ref\cite{17}).
Finally, we show that the event horizon with the static observers in
the Hawking effect can be viewed as the special case in our
discussion, if $X_{0}$ vanishes. The above four properties are
essential and can be easily obtained. In the following section, we
will make more discussions about another interesting property of
these observers-detecting radiation.

\section{Radiation from the null hypersurface}

In the above section, we have showed that the null hypersurface is
just the horizon for its corresponding observers. Now we want to see
whether these observers can detect the radiation or the thermal
spectrum from the null hypersurace.

For the sake of simplicity, we only consider the real scalar field,
and the Klein-Gordon equation in curved space-time is

\begin{equation}
\frac{1}{\sqrt{-g}}\frac{\partial }{\partial X^{\mu }}(\sqrt{-g}g^{\mu \nu }%
\frac{\partial \Psi }{\partial X^{\nu }})-\mu ^{2}\Psi =0.  \label{14}
\end{equation}%
Substitute the metric (6) into (14), we obtain

\begin{equation}
\lbrack -\frac{\partial }{\partial \eta }(r^{2}\frac{\partial \Psi }{%
\partial \eta })+\frac{\partial }{\partial \xi }(r^{2}\frac{\partial \Psi }{%
\partial \xi })]\cdot \frac{1}{g_{11}}+\frac{1}{\sin \theta }\frac{\partial
}{\partial \theta }(\sin \theta \frac{\partial \Psi }{\partial \theta })+%
\frac{1}{\sin ^{2}\theta }\frac{\partial ^{2}\Psi }{\partial \varphi ^{2}}%
=\mu ^{2}r^{2}\Psi .  \label{15}
\end{equation}

Usually, the scalar function $\Psi$ fulfilling the covariant
Klein-Gordon equation in the given spherically symmetrical metric
can always be separated as

\begin{equation}
\Psi (\eta ,\xi ,\theta ,\varphi )=R(\eta ,\xi )Y_{lm}(\theta ,\varphi ).
\label{16}
\end{equation}%
where $Y_{lm}(\theta ,\varphi )$ is the usual spherical harmonics.
Thus equation (15) can be simplified to a radial equation

\begin{equation}
-\frac{\partial }{\partial \eta }(r^{2}\frac{\partial R}{\partial \eta })+%
\frac{\partial }{\partial \xi }(r^{2}\frac{\partial R}{\partial \xi })=[\mu
^{2}r^{2}R-l(l+1)R]g_{11}.  \label{17}
\end{equation}

In the usual quantum field theory, we can define the positive
frequency with respect to $\eta$ by solving equation (17) and then
define the corresponding vacuum state. However, $(\frac{\partial
}{\partial \eta})^{a}$ is not a killing vector, thus the vacuum
state is not invariant, which leads to the fact that the following
discussion about radiation from the null hypersurface or horizon is
just local and only suitable for observers near the null
hypersurface\cite{19}.

Consider the asymptotic expression for the field near the horizon
, and from (7) we obtain

\begin{equation}
\frac{\partial r}{\partial \xi }=\frac{2ae^{a\xi }X_{0}ch(a\eta )+2ae^{2a\xi
}}{\frac{r}{(2M)^{2}}e^{\frac{r}{2M}}},\frac{\partial r}{\partial \eta }=%
\frac{2ae^{a\xi }X_{0}ch(a\eta )}{\frac{r}{(2M)^{2}}e^{\frac{r}{2M}}}.
\label{18}
\end{equation}%
Thus, when $\xi =-\infty $, the location of the event horizon, (17) will be

\begin{equation}
-\frac{\partial ^{2}R}{\partial \eta ^{2}}+\frac{\partial ^{2}R}{\partial
\xi ^{2}}+2r[\frac{\partial R}{\partial \xi }-\frac{\partial R}{\partial
\eta }]\frac{\partial r}{\partial \eta }=0.  \label{19}
\end{equation}%
Obviously, there are two exact solutions

\begin{equation}
R_{\omega }^{in}=e^{-i\omega (\eta +\xi )},R_{\omega }^{out}=\frac{C}{r}%
e^{-i\omega (\eta -\xi )}\text{ \ (here C is the normalized factor)}.
\label{20}
\end{equation}
which are just the two explicit asymptotic expression for the
field near the horizon and can be viewed as the in-going wave and
out-going wave. Because the radiation is local, in what follows
we will use the Damour-Ruffini method to discuss the radiation from the horizon %
\cite{13,20,21,22,23}, although there are many other approaches to
prove the radiation \cite{4,11,14}.

In order to clearly find out the analytic properties of the
out-going wave and in-going wave, we use the advanced
Eddington-Finkelstein coordinates which is defined by $\nu =\eta
+\xi $. And (20) becomes

\begin{equation}
R_{\omega }^{in}=e^{-i\omega \nu },\label{21}
\end{equation}%
\begin{equation}
R_{\omega }^{out}=\frac{C}{r}e^{2i\omega \xi }e^{-i\omega \nu }.
\label{22}
\end{equation}%
From (21) and (22), it's easy to see that the out-going wave is
not analytic in the horizon, while the in-going wave is.

Physically, the analytic in-going wave can be viewed as a
classical particle locally detected by the observers near the
horizon drops into the horizon to reach a negative-energy state.
In the quantum description, this phenomenon allows an antiparticle
to reach positive-energy states from the horizon by the tunneling
effect. Thus, according to Refs\cite{13,18}, the prescription
$r\rightarrow r-i0$ will yield the unique continuation of Eq.(21)
describing an antiparticle state. And it's equivalent to
analytical extension of the out-going wave through the lower
complex plane. To get more insight of the analytical extension, we
can introduce a new coordinates transformation

\begin{equation}
\xi =\frac{1}{2a}\ln \rho .  \label{23}
\end{equation}%

This transformation's most advantage is that it will change the
location of horizon from $\xi =-\infty $ to $\rho =0$. And it is
analogy with the tortoise transformation in the Rindler
spacetime\cite{19}. After taking the transformation, the
$R_{\omega }^{out}$ is

\begin{equation}
R_{\omega }^{out}=\frac{C}{r}\rho ^{i\omega /a}e^{-i\omega \nu }\text{ \ }%
(r>r_{H}).  \label{24}
\end{equation}%
and the metric in (6) becomes

\begin{equation}
ds^{2}=\frac{32M^{3}}{r}e^{-\frac{r}{2M}}\rho a^{2}(-d\eta ^{2}+d\xi
^{2})+r^{2}(d\theta ^{2}+\sin ^{2}\theta d\varphi ^{2}). \label{25}
\end{equation}%
From (24) and (25), it is easy to see that if we exchange $ \rho
\rightarrow \left\vert \rho \right\vert e^{-i\pi} $ through the
lower complex plane, the analytic solution of
$\overset{\symbol{126}}{R}_{\omega }^{out}$ inside the horizon can
be found to be

\begin{equation}
\overset{\symbol{126}}{R}_{\omega }^{out}=\frac{C}{r}e^{\pi \omega
/a}e^{2i\omega \xi }e^{-i\omega \nu }\text{ \ }(r<r_{H}).
\label{26}
\end{equation}%

By introducing the Heaviside function $Y$,

\begin{equation}
\overset{\wedge }{R}_{\omega }^{out}=N_{\omega }[Y(r-r_{H})R_{\omega
}^{out}+Y(r_{H}-r)\overset{\symbol{126}}{R}_{\omega }^{out}].
\label{27}
\end{equation}%
where $N_{\omega }$ is a normalization factor. Eq.(27)is just the
unique continuation of Eq.(21) describing an antiparticle state,
and now it can also describe the splitting of $\widehat{R}_{\omega
}^{out}$ in a wave outgoing from the horizon and a wave falling
into the horizon \cite{13}. Using the relation for the Bosen
particles
%\[\widehat{R}_{\omega }^{out}\]
\begin{equation}
<\overset{\wedge }{R}_{\omega _{1}}^{out},\overset{\wedge
}{R}_{\omega _{2}}^{out}>=-\delta (\omega _{1}-\omega _{2})
\label{28}
\end{equation}%
we obtain

\begin{equation}
N_{\omega }^{2}=(e^{2\pi \omega /a}-1)^{-1}  \label{29}
\end{equation}%
Therefore, the temperature is

\begin{equation}
T=\frac{a}{2\pi }  \label{30}
\end{equation}%
which implicates that there is true radiation locally detected by
the observers near the horizon. And the spectrum is also Planckian
, though the effective temperature is obviously low which is the
same as many other thermal effects by radiation. Note that $a$ is
a constant correlated with the proper acceleration of the observer
\cite {19}.

\section{Conclusion and discussion}

In the above sections, we have defined some accelerating observers
by defining the null hypersurface in the Kruskal coordinates at
first. Some properties of these observers such as their 4-velocity,
4-acceleration and particularly the radiation effect are also
obtained. However, as mentioned in the title, what we want to
discuss are the properties of accelerating observers in the
Schwarzschild space. Thus, we should first check that the observers
we defined are truly accelerating in the Schwarzschild space.

In section 2, we have proved that the observers defined in (4) are
accelerating observers in the Schwarzschild spacetime. Considering
the constant proper acceleration of the static observers in the
Schwarzschild space, we can conclude qualitatively that the
observers we defined must be accelerating observers in the
Schwarzschild space. In fact, if we use the Schwarzschild
coordinates $\{t,r,\theta ,\varphi \}$ to express the worldline in
(4), we can obtain

\begin{equation}
(\frac{r}{2M}-1)^{1/2}e^{\frac{r}{4M}}=X_{0}ch\frac{t}{4M}+\sqrt{X_{0}^{2}sh%
\frac{t}{4M}+e^{2a\xi _{0}}}.  \label{30}
\end{equation}%
which shows that the observer defined in (4) is truly an observer
accelerated in the space. And it can be viewed as a detector who is
at first accelerated closely to the planet, and then gets far away
from it (Figure 2). It's obvious to find that as the same as that in
many other similar works, the detected effective temperature in (29)
is low to detect in the experiment. However, the theoretical
discussion is still of sense, which can be seen in the following.

Property of detecting the radiation for accelerating observers is
seemingly not surprising due to the analogy of Unruh effect.
However, these observers accelerate in the space, not the spacetime,
this may make some difference for the result. For example, though
the free drop observer accelerates in the Schwarzschild space, it
will detect zero radiation due to the fact that it's a geodesic
observer in the spacetime, while the static observer in the
Schwarzschild space will detect the well-known Hawking effect. Of
course, the key point of deciding whether there is radiation is due
to the 4-acceleration of the observer. However, according to our
proof, the more significant key point is whether there exists the
corresponding null hypersurface which can be treated as the horizon
for the observers. It's true that, if there is a null hypersurface
we can always find the corresponding observers, while the opposite
argument may not be correct. Thus, we conjecture that, for any null
hypersurface in any spacetime, we can find the corresponding
observers who can detect (at least locally) the radiation from it.
In fact, in Ref\cite{15}, the authors used the similar trick,
defining the null hypersurface at first, and then discussed the
properties of some accelerating observers in Minkowski spacetime.
Their results can be viewed as a partial support to our conjecture.
Particularly, in Ref\cite{16}, the authors qualitatively obtained
the result of the conjecture from the equivalent principle of
General Relativity. Moreover, they applied it to define the local
temperature and obtained very interesting results.

\section{Acknowledgements}

Ya-Peng Hu thanks Dr.Xuefei Gong and the referees for useful
discussions and help. And Ya-Peng Hu is supported partially by
grants from NSFC, China (No. 10325525 and No. 90403029), and a grant
from the Chinese Academy of Sciences. Zheng Zhao is supported by the
National Natural Science Foundation of China (Grant No. 10773002)
and the National Basic Research Program of China (Grant No.
2003CB716302).

\begin{figure}[tbh]
\includegraphics[width=0.6\textwidth]{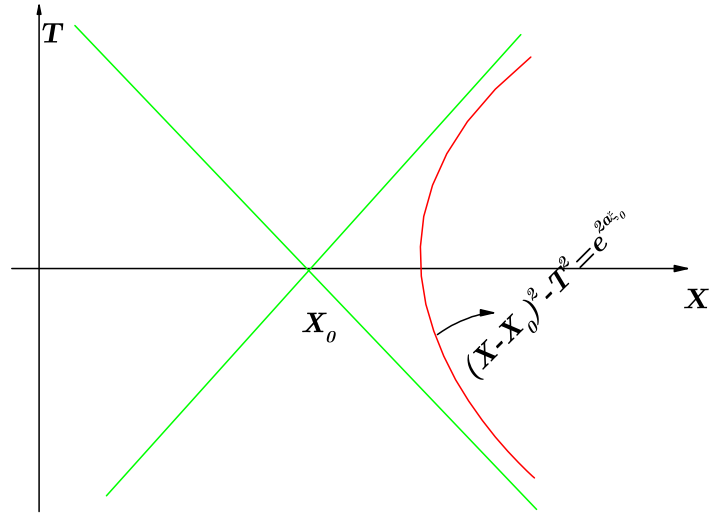}
\caption{(Color online) Null hypersurface and the corresponding
observer by definition.}
\label{Fig.1}
\end{figure}

\begin{figure}[tbh]
\includegraphics[width=0.6\textwidth]{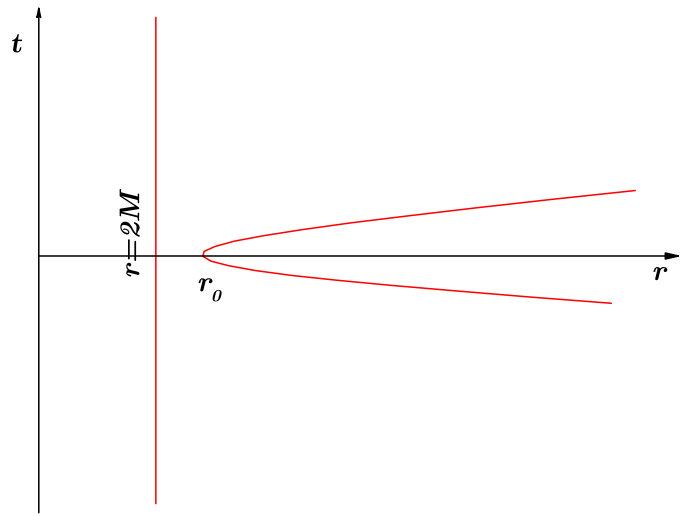}
\caption{(Color online) The behavior of the observer in the
Schwarzschild coordinates.}
\label{Fig.2}
\end{figure}

\end{document}